# A short gamma-ray burst apparently associated with an elliptical galaxy at redshift z=0.225

N. Gehrels[1], C.L. Sarazin[2], P.T. O'Brien[3], B. Zhang[4], L. Barbier[1], S.D. Barthelmy[1], A. Blustin[5], D.N. Burrows[6], J. Cannizzo[1, 7], J.R. Cummings[1, 8], M. Goad[3], S.T. Holland[1, 9], C.P. Hurkett[3], J.A. Kennea[6], A. Levan[3], C.B. Markwardt[1, 10], K. O. Mason[5], P. Meszaros[6], M. Page[5], D.M. Palmer[11], E. Rol[3], T. Sakamoto[1, 8], R. Willingale[3], L. Angelini[1, 7], A. Beardmore[3], P.T. Boyd[1, 7], A. Breeveld[5], S. Campana[12], M.M. Chester[6], G. Chincarini[12, 13], L.R. Cominsky[14], G. Cusumano[15], M. de Pasquale[5], E.E. Fenimore[11], P. Giommi[16], C. Gronwall[6], D. Grupe[6], J.E. Hill[6], D. Hinshaw[1, 17], J. Hjorth[18], D. Hullinger[1, 10], K.C. Hurley[19], S. Klose[20], S. Kobayashi[6], C. Kouveliotou[21], H.A. Krimm[1, 9], V. Mangano[12], F.E. Marshall[1], K. McGowan[5], A, Moretti[12], R.F. Mushotzky[1], K. Nakazawa[22], J.P. Norris[1], J.A. Nousek[6], J.P. Osborne[3], K. Page[3], A.M. Parsons[1], S. Patel[23], M. Perri[16], T. Poole[2], P. Romano[12], P.W.A. Roming[6], S. Rosen[5], G. Sato[22], P. Schady[5], A.P. Smale[24], J. Sollerman[25], R. Starling[26], M. Still[1,9], M. Suzuki[27], G. Tagliaferri[12], T. Takahashi[22], M. Tashiro[27], J. Tueller[1], A.A. Wells[3], N.E. White[1], & R.A.M.J. Wijers[26]

1. *NASA/Goddard Space Flight Center Greenbelt, Maryland 20771, USA*
2. *Department of Astronomy, University of Virginia, Charlottesville, Virginia 22903-0818, USA*
3. *Department of Physics & Astronomy, University of Leicester, Leicester,LE1 7RH, UK*

*4. Department of Physics, University of Nevada, Las Vegas, Las Vegas, Nevada 89154-4002, USA*

*5. Mullard Space Science Laboratory, University College London, Dorking, UK RH5 6NT*

*6. Department of Astronomy and Astrophysics, Penn State University, University Park, PA 16802, USA*

*7. Joint Center for Astrophysics, University of Maryland, Baltimore County, Baltimore, Maryland 21250, USA*

*8. National Research Council, 2101 Constitution Ave NW, Washington, DC 20418, USA*

*9. Universities Space Research Association, 10211 Wincopin Circle, Suite 500, Columbia, Maryland 21044-3432, USA*

*10. Department of Astronomy, University of Maryland, College Park, Maryland 20742, USA*

*11. Los Alamos National Laboratory, Los Alamos, New Mexico 87545, USA*

*12. INAF - Osservatorio Astronomico di Brera, Via Bianchi 46, I-23807 Merate, Italy*

*13. Universita degli studi di Milano Bicocca, P.za delle Scienze 3, I-20126 Milano, Italy*

*14. Department of Physics and Astronomy, Sonoma State University, Rohnert Park, California 94928, USA*

*15. INAF - Istituto di Astrofisica Spaziale e Cosmica, Via Ugo La Malfa 153, I-90146 Palermo, Italy*

*16. ASI Science Data Center, Via Galileo Galilei, I-00044 Frascati, Italy*

*17. SP Systems Inc., 7500 Greenway Center Dr., Greenbelt, Maryland 20770, USA*

*18. Niels Bohr Institute, University of Copenhagen, DK-2100 Copenhagen, Denmark*

*19. UC Berkeley Space Sciences Laboratory, Berkeley, California 94720-7450. USA*

*20. Thüringer Landessternwarte Tautenburg, Sternwarte 5, D-07778 Tautenburg, Germany*

*21. NASA/Marshall Space Flight Center, NSSTC, XD-12, 320 Sparkman Drive, Huntsville, Alabama 35805, USA*

*22. Institute of Space and Astronautical Science,JAXA, Kanagawa 229-8510, Japan*

*23. Universities Space Research Association, NSSTC, XD-12, 320 Sparkman Drive, Huntsville, Alabama 35805, USA*

*24. Office of Space Science, NASA Headquarters, Washington, DC 20546, USA*

*25. Stockholm Observatory, Department of Astronomy, AlbaNova, 106 91 Stockholm, Sweden*

*26. Astronomical Institute "Anton Pannekoek", University of Amsterdam, Kruislaan 403, 1098 SJ Amsterdam, The Netherlands*

*27. Department of Physics, Saitama University, Sakura, Saitama, 338-8570, Japan*

**Gamma Ray Bursts (GRBs) are bright, brief flashes of high energy photons that have fascinated scientists for 30 years. They come in two classes[1]: long (>2 s), soft-spectrum bursts and short, hard events. The major progress to date on understanding GRBs has been for long bursts which are typically at high redshift (z ~ 1) and are in sub-luminous star-forming host galaxies. They are likely produced in core-collapse explosions of massive stars[2]. Until the present observation, no short GRB had been accurately (<10") and rapidly (minutes) located. Here we report the detection of X-ray afterglow from and the localization**

**of short burst GRB050509b. Its position on the sky is near a luminous, non-star-forming elliptical galaxy at a redshift of 0.225, exactly the type of location one would expect if the origin of this GRB is the long-proposed[3,4] fiery merger of neutron star (NS) or black hole (BH) binaries. The X-ray afterglow is found to be weak and fading below detection within a few hours and no optical afterglow is detected to stringent limits, explaining the past difficulty in localizing short GRBs.**

The new observations are from the *Swift*[5] satellite which features the hard X-ray wide-field Burst Alert Telescope (BAT), and rapid spacecraft slewing to point the narrow-field X-Ray Telescope (XRT) and UV-Optical Telescope (UVOT) at the burst. On 2005 May 9 at 04:00:19.23 UT, the BAT triggered and located GRB050509B on-board[6]. The BAT location is shown in Fig. 1 and light curves in Fig. 2. The event is a single short spike with duration of 40±4 ms. The burst has a ratio of 50-100 keV to 25-25 keV fluences of 1.4±0.5 which is consistent with, but in the soft portion of, the short/hard population detected by the first extensive GRB survey made with the Burst and Transient Source Experiment (BATSE). The 15-150 keV fluence is $(9.5\pm2.5)\times10^{-9}$ erg $cm^{-2}$, which is the lowest imaged by BAT to date and is just below the short GRB fluence range detected by BATSE (adjusted for different energy ranges of the two instruments).

*Swift* slewed promptly and XRT started acquiring data 62 s after the burst (T+62 s, where T is the BAT trigger time). Ground processed data revealed an uncataloged X-ray source near the center of the BAT error circle containing 11 photons (5.7σ significance due to near-zero background in image) in the first 1640 s of integration time. The XRT position is shown with respect to the Digitized Sky Survey (DSS) field in Fig. 1. A *Chandra* Target of Opportunity observation of the XRT error circle was performed on May 11 at 4:00 UT for 50 ks, with no sources detected in the XRT error circle. The light curve combining BAT, XRT, and *Chandra* data is shown in Fig. 3. The UVOT observed the field starting at T+60 s. No new optical/UV sources were found in the XRT error circle to V-band magnitude >19.7 for t<300 min.

*Swift* has provided the first accurate localization of a short GRB. No optical afterglow was detected to stringent limits (R-band magnitude >25 at 25 hrs[7]). When the XRT error circle is plotted on the R-band image we obtained[8] with the Very Large Telescope (VLT), several faint objects are seen in the error circle, some of which are extended and could be high redshift galaxies[9,10]. It is possible the burst occurred in one of these. However, the center of the XRT error circle lies only 9.8" away from the center of the large E1 elliptical galaxy 2MASX J12361286+2858580[10] at a redshift of 0.225[11], which is located in the cluster NSC J123610+285901[12,13]. This is a luminous giant elliptical galaxy; its 2 Micron All Sky Survey (2MASS) magnitude of K = 14.1 corresponds to a luminosity of $4\times10^{11}$ $L_{O} \approx 3$ $L^{*}$, assuming standard cosmology. Our *Chandra* image shows that this is the central dominant galaxy in one of two merging subclusters in this bimodal cluster. Although caution is always prudent for *a posteriori* statistics, the association with this galaxy seems unlikely to be coincidental. The probability of a random location being within 10" of a galaxy with an apparent magnitude at least this bright is $\sim10^{-3}$. Moreover, galaxies this luminous are relatively rare; the comoving number density[14] of galaxies at least this luminous is $\sim 5\times10^{-5}$ $Mpc^{-3}$; the probability of lying within 10" of a randomly located one at $z \leq 0.225$ is $\sim10^{-4}$. Note that this is the first GRB out of ~80 with accurate optical localizations to be near a bright elliptical on the sky.

The likely association between GRB050509B and 2MASX J12361286+2858580 is difficult to understand if the GRB resulted from any mechanism involving recent star formation. The galaxy type for the suggested host galaxy is very different than those found for long GRBs; their hosts are typically sub-luminous and blue[15] and show strong emission lines associated with star formation[16]. As is true of most giant ellipticals in clusters, 2MASX J12361286+2858580 has no indications of UV or optical line emission[10]. Our UVOT images clearly detect the galaxy in the optical, but not in the UV (UVM2 220 nm and UVW2 188 nm filters), as expected for an elliptical galaxy implying little or no contribution from young, hot stars. The 3σ upper limit at 188 nm gives a limit on the star formation rate[17] of < 0.2 $M_O$ $yr^{-1}$. It would be improbable to find a massive-star core collapse or young magnetar in this galaxy. In addition, the isotropic energy of $1.1x10^{48}$ k erg (15-150 keV, z=0.225, k is k-correction factor which is typically 1 to 10) is $>10^2$ times higher than that of the 2004 Dec 27 giant flare from SGR 1806-20[18,19]. Thus, it is unlikely that this burst was an SGR-type flare.

On the other hand, 2MASX J12361286+2858580 is a very propitious site for a NS-NS or NS-BH merger. As *Chandra* observations have shown[20], giant ellipticals, especially those dominant in their cluster, have large populations of low mass X-ray binaries (LMXBs) containing accreting NSs and BHs. Further, a high fraction (>~50%) of the LMXBs in ellipticals are located in globular clusters (GCs)[21] because close binary systems containing at least one compact object can easily be formed dynamically in GCs. While there is less direct evidence that close double NS binaries can form easily in GCs, the double NS system PSR B2127+11C in the Galactic GC M15 is an example of such a binary[22], and has a merger lifetime of ~ $2x10^8$ yr. In fact, of all galaxy types, large ellipticals (particularly cluster dominants) are the most likely place to find double compact binary systems due to their large populations of GCs.

The center of the XRT error circle is in the outer regions of the elliptical galaxy, although the circle extends nearly to the galaxy center. A location at large radius would be consistent with the binary merger model for short GRBs[23]. NS-NS binaries often obtain significant kick velocities (100-1000 km $s^{-1}$) from the supernova that creates each NS[24]. If one ignores the effects of the galactic potential, a NS-NS binary moving at 1000 km $s^{-1}$ would travel 100 kpc in $10^8$ yr. The projected distance of the center of the XRT position from the center of the galaxy is about 35 kpc; the range over the error circle is about 2-70 kpc. Thus, the NS-NS binary might have reached this distance prior to merging, even if it started from a more central location. Alternatively, the binary may have formed in a GC; GCs have a broader radial distribution than field stars in ellipticals, which could explain the large projected radius of the GRB.

The X-ray emission for GRB050509B is faint, being the weakest afterglow of any of the 15 GRBs that XRT has promptly observed (factor ~200 weaker than XRT average). For BATSE bursts, studies were done of the post-burst emission by summing large numbers of GRB lightcurves[25,26]. The post-burst emission was found to be weaker for short bursts than for long events, consistent with the GRB050509B. For typical shock parameters, the early X-ray afterglow emission is likely below the cooling frequency[27]; in this regime, the weak afterglow is consistent with the low density medium around an evolved compact binary progenitor. A more critical factor to define the low X-ray flux may be the small energy injection involved, as the prompt emission for GRB050509B is also the weakest of the BAT GRBs. If the redshift is 0.225, then the afterglow is >100 times less luminous than that of typical long-burst afterglows and the isotropic energy is ~$10^{-4}$ that of typical long GRBs (about the same as the lowest luminosity, unusual GRB 980425).

Prior to *Swift*, it was predicted[28] that short GRBs would have faint optical afterglows, particularly so if occurring in low density regions like those around evolved stars. This prediction is consistent with the lack of optical detection to stringent limits for GRB050509B, although one should keep in mind that this burst is weak compared to other short GRBs. It is likely that X-ray afterglow will remain a key to understanding short bursts .

The X-ray afterglow from this short GRB can constrain outflow parameters. The fact that the X-rays are fading as early as 62 s puts a limit on the initial Lorentz factor of $\Gamma_0 \geq 70\ n_{-2}^{-1/8} E^{1/8}$ for z=0.225 ($n_{-2}$ is ambient density in units of $10^{-2}$ $cm^{-3}$, E is the isotropic energy in units of $10^{48}$ erg), showing that short GRBs are highly relativistic events.

Another interesting aspect of the localization of GRB050509B is that the burst is faint and yet has a bright galaxy in its error circle. There are 5 previous short GRBs with fluences 1 to 3 orders of magnitude larger than GRB050509B that have had their arcmin-sized error boxes searched for bright galaxies[29,30]. There are galaxies in each error box of brightness comparable to or less than 2MASX J12361286+2858580, but none much brighter as one might expect for these brighter GRBs. This does not contradict a merger model for short GRBs, since, while giant elliptical galaxies are a rich environment for mergers, most would occur in the more numerous, fainter, star-forming galaxies. Thus star-forming galaxies harbor both massive stars and evolved binaries, while ellipticals have almost no star formation and are highly deficient in short-lived massive stars. The detection of GRB050509B near an elliptical galaxy is an important observation for short bursts since the association with a large elliptical galaxy is evidence against a collapsar origin, whereas an association with a star-forming galaxy would have left the question unanswered. There may be more than one origin of short GRBs, but this particular short event has a high probability of being unrelated to star formation and caused by a binary merger.

---

**Acknowledgements** The authors acknowledge support from ASI, NASA and PPARC.

**Competing interests statement** The authors declare that they have no competing financial interest.

**Correspondence** should be addressed to Neil Gehrels at gehrels@milkyway.gsfc.nasa.gov

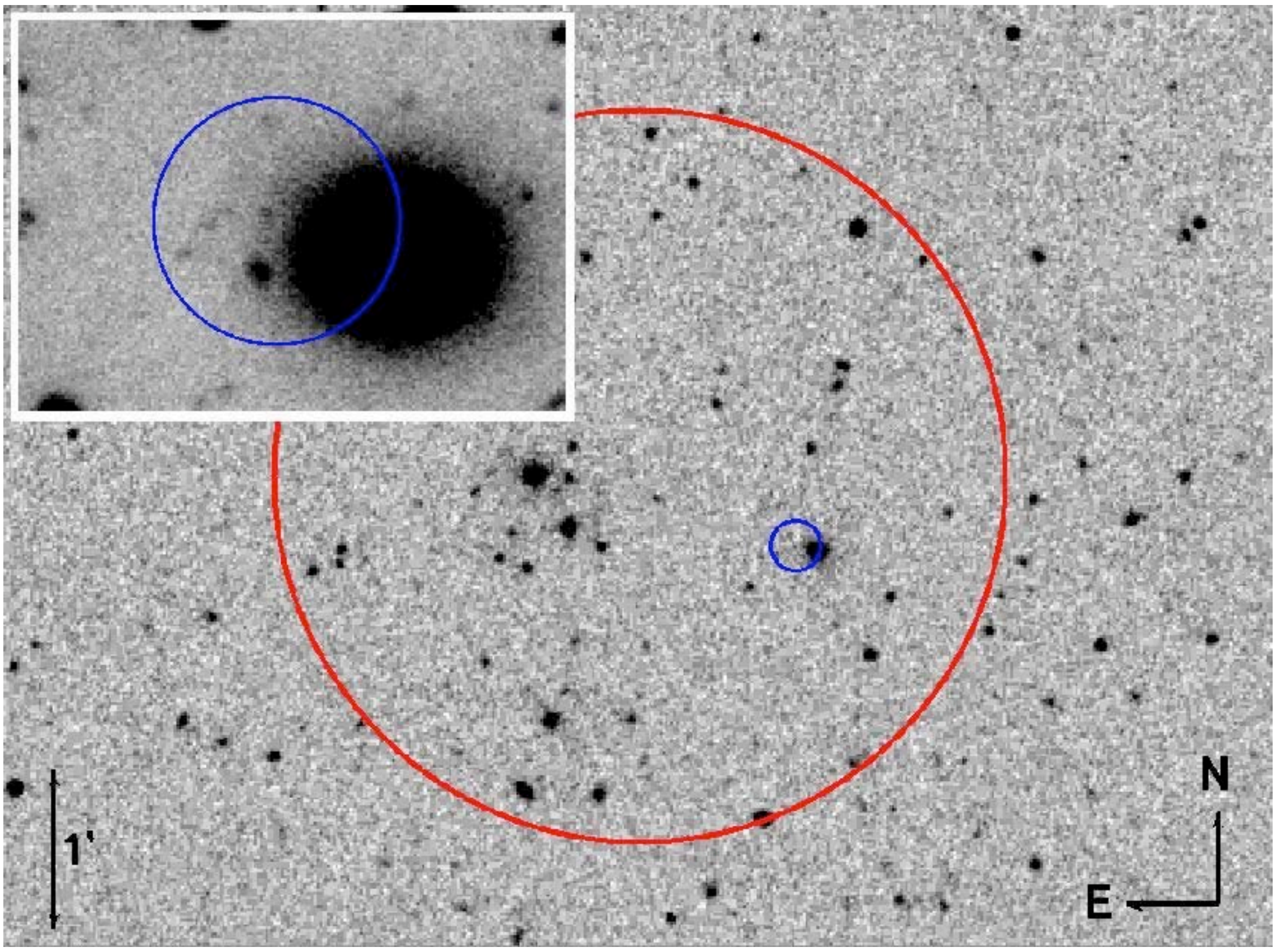

**Figure 1** Optical images of the region of GRB050509B showing the association with a large elliptical galaxy. Large: Digitized Sky Survey (red) image. Large red circle is the BAT position error circle, while the smaller solid blue circle is the XRT position error circle. The BAT position is 12h 36m 18s, +28° 59' 28" (J2000) with 2.3' error radius (90% containment). The XRT, operating in its most sensitive Photon Counting mode, derived a position of 12h 36m 13.58s, 28° 59' 01.3" (J2000), with a positional accuracy of 9.3" (90% containment radius; larger than typical XRT 4" accuracy due to weakness of burst). This position takes into account the low counting statistics, cluster emission in the field and astrometric corrections[10] to the 2MASS coordinate system.

Many of the extended objects are likely to be galaxies in the cluster NSC J123610+28590131. Inset: Blow-up of the region of the XRT error circle from an R-band image obtained[8] using FORS2 on the 8.2m VLT-Antu telescope at the European Southern Observatory/Paranal on May 11.0 UT, 1.85 days after the burst. The extended source to the right (west) is the luminous elliptical galaxy 2MASX J12361286+285858026, which we suggest as the likely host of the burst. Other objects in the error circle are not identified, but appear to be faint galaxies either associated with the same cluster as the elliptical galaxy or at higher redshift. The VLT image consists of fifteen 3-min frames taken under good conditions (~1" seeing).

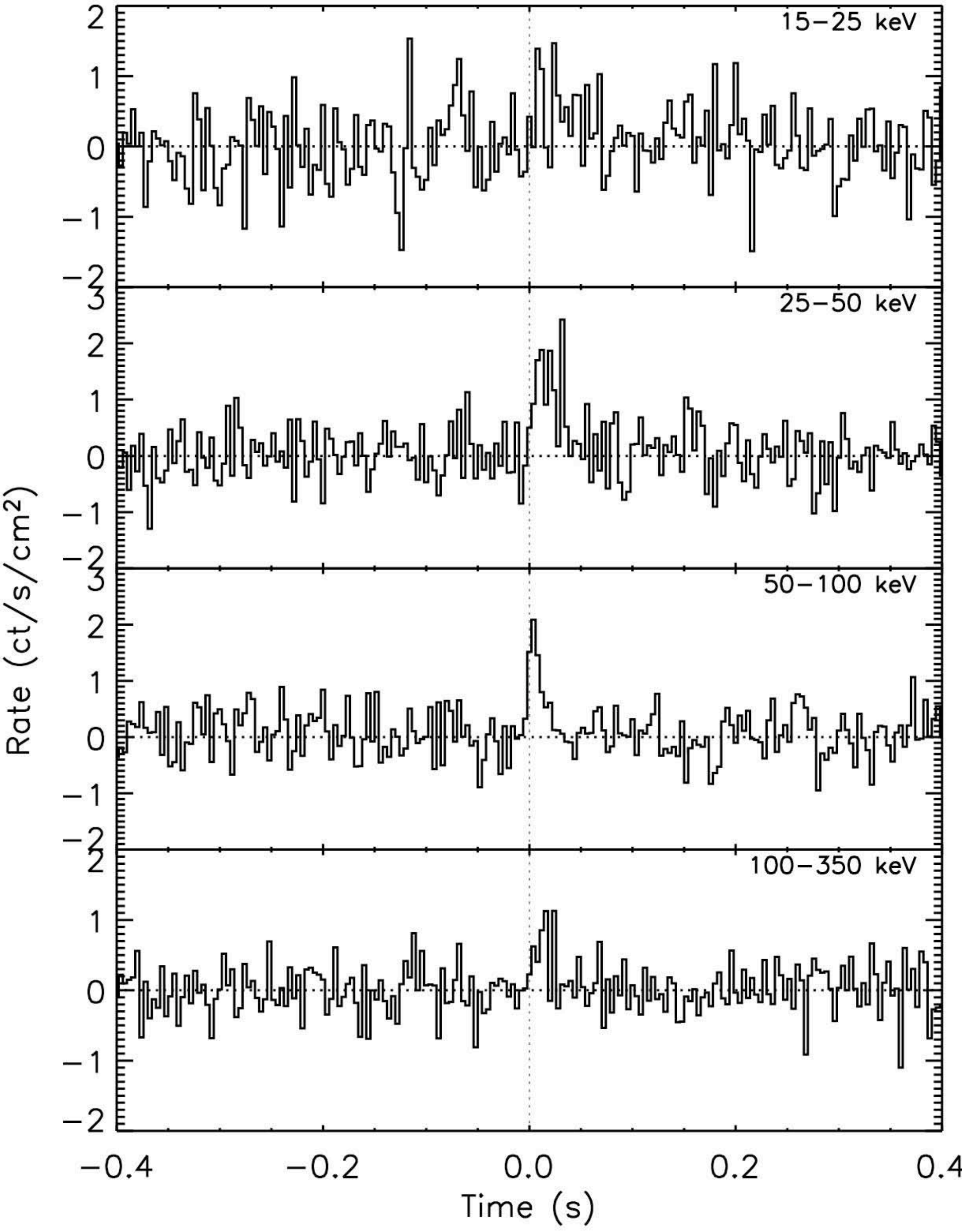

**Figure 2** BAT light curves for the short GRB050509B showing the short duration of this GRB. The light curves are given in four photon energy bands with the band identified in the upper right of each panel. The peak has a duration of 40±4 ms (90% containment of counts). There is no detectable emission except from T-30 ms to T+30 ms, confirming the "short" aspect of this burst. The successful trigger criterion for the GRB was in the 25-100 keV band. The peak count rate measured by BAT is ~2100 c $s^{-1}$ in the 15-150 keV band at T+5 ms. The BAT data (40ms of data centered on T+23 ms) are well fit by a simple power law model with a photon index of 1.5±0.4, a normalization at 50 keV of (2.0±0.5)x$10^{-2}$ photons $cm^{-2}$ $s^{-1}$ $keV^{-1}$ and a peak flux of 2.53±0.33 ph $cm^{-2}$ $s^{-1}$ (all in 15-150 keV and 90% confidence level).

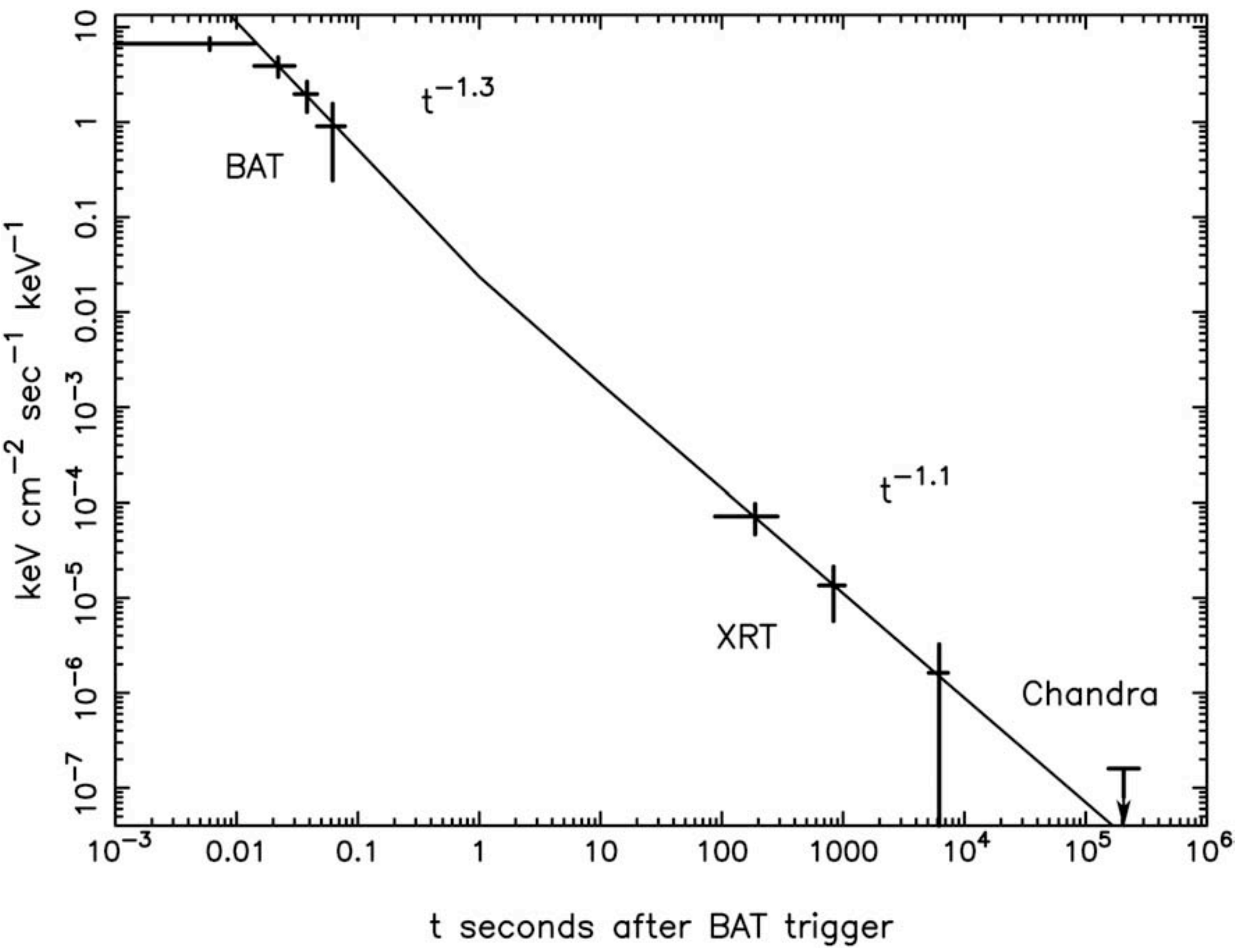

**Figure 3** X-ray afterglow light curve for GRB050509B showing weak flux falling off to undetectability after $10^4$ seconds. The decline in the 1 keV X-ray emission from GRB050509B from BAT, XRT, and Chandra data is shown. The BAT points were calculated by extrapolating the BAT spectrum with measured photon index $\Gamma$ (flux = constant x $E^{-\Gamma}$) of 1.5 down to 1 keV. The XRT fluxes are unabsorbed values derived by fixing the photon index to 1.5 (i.e., that found by the BAT) and using the Galactic column density of $1.5 \times 10^{20}$ $cm^{-2}$. The initial flux is $(3.57 \pm 1.08) \times 10^{-13}$ erg $cm^{-2}$ $s^{-1}$ between 0.3 and 10 keV. The XRT time decay index $\alpha$ (flux = constant x $time^{-\alpha}$) is 1.10 (0.57 to 2.36 90% conf) and the BAT index is 1.34 (0.27 to 2.87 90% conf). The BAT index is poorly constrained because the burst is so short. The BAT and XRT data are also consistent at the 90% confidence level with a single decay index of 1.20 (1.12 to 1.29 90% conf). The Chandra 99% confidence flux upper limit lies above the decay curves.